\documentclass[twocolumn,twoside,preprintnumbers,superscriptaddress,nofootinbib,natbib,bm,amsrefs]{revtex4-2}

\usepackage{multirow,array}
\usepackage{amsmath,amssymb}
\usepackage{graphicx}
\usepackage{color}
\usepackage{slashed}
\usepackage{footmisc}
\usepackage{comment}
\usepackage{braket}
\usepackage{colortbl}
\usepackage{orcidlink}
\usepackage{physics}
\usepackage{url}
\usepackage{placeins}

\newcommand{\whizard}{\texttt{Whizard}}

\newcommand{\ov}{\overline}
\newcommand{\cba}{c_{\beta\alpha}}
\newcommand{\tb}{t_{\beta}}
\newcommand{\caosb}{c_{\alpha}/s_{\beta}}

\definecolor{nicered}{rgb}{0.5,0.,0.}
\definecolor{nicegreen}{rgb}{0.,0.5,0.}
\definecolor{niceblue}{rgb}{0.,0.,0.5}

\usepackage{hyperref}

\hypersetup{colorlinks=true,citecolor=nicegreen,linkcolor=nicered,urlcolor=kitgreen}

\definecolor{kitgreen}{rgb}{0,
0.58823 
, 0.50980 
}

\allowdisplaybreaks

\begin{document}

\title{Chasing the two-Higgs-doublet model via electroweak corrections at $e^+ e^-$ colliders}

\author{Pia Bredt
\orcidlink{0000-0003-4579-0387}
} 
\affiliation{Center for Particle Physics Siegen, University of Siegen, 
Walter-Flex-Str. 3, 
57072 Siegen, Germany}

\author{Tatsuya Banno
\orcidlink{0009-0009-1026-2534}
}
\affiliation{Department of Physics,
Nagoya University, Nagoya 464-8602, Japan}

\author{Marius H\"ofer
\orcidlink{0000-0003-0009-9410}
} 
\affiliation{Institute for Theoretical Physics, Karlsruhe Institute of Technology, 
Wolfgang-Gaede-Str.\,1, 
76131 Karlsruhe, Germany}

\author{Syuhei Iguro
\orcidlink{0000-0002-8295-9623}
} 
\affiliation{Institute for Advanced Research, Nagoya University, Nagoya 464--8601, Japan}
\affiliation{Kobayashi-Maskawa Institute for the Origin of Particles and the Universe, Nagoya University, Nagoya 464--8602, Japan}

\author{Wolfgang Kilian
\orcidlink{0000-0001-5521-5277}
} 
\affiliation{Center for Particle Physics Siegen, University of Siegen, 
Walter-Flex-Str. 3, 
57072 Siegen, Germany}

\author{Yang Ma
\orcidlink{0000-0002-9419-6598}
} 
\affiliation{Center for Cosmology, Particle Physics and Phenomenology, Universit\'e catholique de Louvain, B-1348 Louvain-la-Neuve, Belgium}

\author{J\"urgen Reuter
\orcidlink{0000-0003-1866-0157}
} 
\affiliation{Deutsches Elektronen-Synchrotron DESY, 
Notkestr. 85, 
22607 Hamburg, Germany}

\author{Hantian Zhang
\orcidlink{0000-0002-7472-6467}
} 
\email{hantian.zhang@cern.ch}
\affiliation{
Theoretical Physics Department, CERN, 1211 Geneva 23, Switzerland
}

\preprint{CERN-TH-2025-181, DESY-25-121, SI-HEP-2025-19, P3H-25-063, IRMP-CP3-25-30, KA-TP-28-2025}

\begin{abstract}
We present a comprehensive study of Higgs boson production associated with a neutrino pair at $e^+e^-$ colliders ($e^+ e^- \to h \, \nu \bar{\nu}$)
at next-to-leading-order accuracy in both the Standard Model and the two-Higgs-doublet model. 
We show that these new physics effects  will be
observable in total and differential cross sections when compared with theoretical predictions that
include electroweak corrections, even in the Higgs alignment limit.
This highlights the potential of precision studies at future $e^+e^-$ colliders for searching new physics.

\end{abstract}

\maketitle

The Higgs boson
discovered at the Large Hadron Collider~(LHC)~\cite{ATLAS:2012yve,CMS:2012qbp} has completed the particle content predicted by the Standard Model~(SM)~\cite{Glashow:1961tr,Weinberg:1967tq,Salam:1968rm}.
Despite its great success, fundamental problems such as the origins of baryon asymmetry of the universe and dark matter persist, necessitating new physics beyond the SM~(BSM).
In this context, precise studies of Higgs boson properties and direct searches of new physics form two complementary focuses in modern particle physics.
Over the past decade, the accuracy of Higgs-sector measurements has been greatly improved at the LHC, while there is still room for new physics effects at the ten percent level~\cite{CMS:2022dwd,ATLAS:2022vkf}. 
Although the LHC is well-suited for studying the gross picture of the Higgs sector,
future $e^+e^-$ colliders such as FCC-ee, CEPC,
ILC/LCF and CLIC~\cite{FCC2018,FCC:2025lpp,CEPC2018,CEPCStudyGroup:2023quu,ILC2022,LinearColliderVision:2025hlt,Linssen:2012hp,Adli:2025swq} are required for the determination of Higgs boson properties at the electroweak scale with a few permille accuracy.

In this paper, we show that there are realistic opportunities for searching new physics through the precision program at future high-energy $e^+e^-$ colliders, particularly via electroweak~(EW) corrections to scattering processes.
We demonstrate this possibility through a comprehensive study of single Higgs production ($e^+ e^- \to h + \nu_{\ell} \bar{\nu}_{\ell}$ with $\ell=e,\mu,\tau$) at next-to-leading-order~(NLO) EW accuracy in both the SM and two-Higgs-doublet model~(2HDM).
The 2HDM is one of the simplest extensions of the SM and has rich phenomenology~\cite{Lee:1973iz}.
This model can resolve the vacuum meta-stability issue~\cite{Sher:1988mj} and provide strong first-order electroweak phase transition for baryogenesis~\cite{Cohen:1993nk} and detectable gravitational waves~\cite{Dorsch:2016nrg}.
It often appears as a low-energy scalar sector of more UV-complete theories, for example, the left-right symmetric models \cite{Pati:1974yy,Mohapatra:1974hk,Senjanovic:1975rk} and supersymmetric~(SUSY) models \cite{Fayet:1977yc,Martin:1997ns}.
In the 2HDM there exists an alignment limit in which the discovered Higgs boson behaves exactly as the SM one at the tree level.
We show that even in this limit NLO EW corrections involving new particles can induce few percent effects with respect to the SM predictions,
thereby offering an important window to access this model.

In the literature, the on-shell Higgsstrahlung process, $e^+ e^- \to Zh$, has been intensively investigated.
For example, full NLO EW corrections in the 2HDM and SUSY models are computed in Refs.~\cite{Xie:2018yiv,Aiko:2021nkb,Anisha:2025hkk,Heinemeyer:2025crt}, 
and the higher-order calculations in the SM for related triangle and box form factors are computed in Refs.~\cite{Kniehl:1995at,Gong:2016jys,Sun:2016bel,Wang:2019fxh,Ma:2021cxg,Wang:2021imw,Freitas:2022hyp}.
However, the more complicated off-shell single Higgs production $e^+ e^- \to h \, \nu \bar{\nu}$ is not so well studied.
Pioneering works for this $2\to 3$ process include the full NLO EW corrections in the SM~\cite{Belanger:2002ik,Denner:2003yg,Denner:2003iy}
and one-loop triangle form factor corrections in the SUSY models~\cite{Eberl:2002xd,Eberl:2002su,Hahn:2002gm,Wang:2006gp},
while a complete NLO EW study in BSM theories is still missing.
This process is advantageous, since at higher center-of-mass energies, $\sqrt{s}=365$ and $550$\,GeV, the cross section is roughly an order of magnitude larger than those of $e^+e^- \to h \, \ell\bar{\ell}$ processes.
In particular, the fact that its cross section at larger $\sqrt{s}$ is dominated by $WW$ fusion allows an independent probe of new physics effects other than $Zh$ production.

Regarding the 2HDM we follow the convention of Refs.~\cite{Aoki:2009ha,Denner:2017vms,Denner:2018opp}.
The two Higgs doublets are denoted by $\Phi_1$ and $\Phi_2$, each acquiring a vacuum expectation value (vev).
If both doublets couple to fermions as in the SM, the neutral scalars induce large tree level flavor changing neutral currents (FCNCs) in general.
To prevent such large FCNCs, it is useful to assign a $\mathbb{Z}_2$~symmetry charge~\cite{Paschos:1976ay,Glashow:1976nt}.
As a result, there are four types of Yukawa structures depending on the $\mathbb{Z}_2$ charge assignment \cite{Barger:1989fj,Grossman:1994jb}.
The CP-conserving Higgs potential is given by
\begin{align}
    & V = \;
    m_1^2 \Phi_1^{\dagger} \Phi_1 + 
    m_2^2 \Phi_2^{\dagger} \Phi_2 
    - m_{12}^2 \big( \Phi_1^{\dagger} \Phi_2 + \Phi_2^{\dagger} \Phi_1 \big) \nonumber \\
    & \; + \frac{\lambda_1}{2}\big( \Phi_1^{\dagger} \Phi_1 \big)^2
    + \frac{\lambda_2}{2}\big( \Phi_2^{\dagger} \Phi_2 \big)^2
    + \frac{\lambda_3}{2}\big( \Phi_1^{\dagger} \Phi_1 \big)\big( \Phi_2^{\dagger} \Phi_2 \big) \nonumber \\
    &\; + \frac{\lambda_4}{2}\big( \Phi_1^{\dagger} \Phi_2 \big)\big( \Phi_2^{\dagger} \Phi_1 \big)  + \frac{\lambda_5}{2}\Big[\big( \Phi_1^{\dagger} \Phi_2 \big)^2 + \big( \Phi_2^{\dagger} \Phi_1 \big)^2 \Big] \,,
\end{align}
where the $\mathbb{Z}_2$ symmetry ($\Phi_1 \to - \Phi_1$ and $\Phi_2 \to \Phi_2$) is softly broken by the $m_{12}^2$ term. 
After spontaneous symmetry breaking, two mixing angles $\alpha$ and $\beta$ can be introduced to obtain the mass eigenstates.
In addition to the 125\,GeV SM-like Higgs boson $h$, we have four scalars: charged scalars~$H^\pm$, neutral scalar~$H$ and pseudoscalar~$A$.
The parameter $c_{\beta\alpha} \equiv \cos(\beta-\alpha)$ governs the mixing between CP-even scalars $h$ and $H$, while $\tb \equiv \tan (\beta)$ represents the ratio of two vevs and determines the Yukawa couplings.
Note that the Higgs alignment limit is realized in $\cba = 0$ such that $h$ is aligned with the SM Higgs. 
Now the extended Higgs sector is parametrized by $\cba,\, \tb,\,\lambda_5,\,m_H,\,m_A,\,m_{H^\pm}$. 

\begin{figure}[t]
    \centering
    \begin{tabular}{ccc}
        \includegraphics[height=1.6cm]{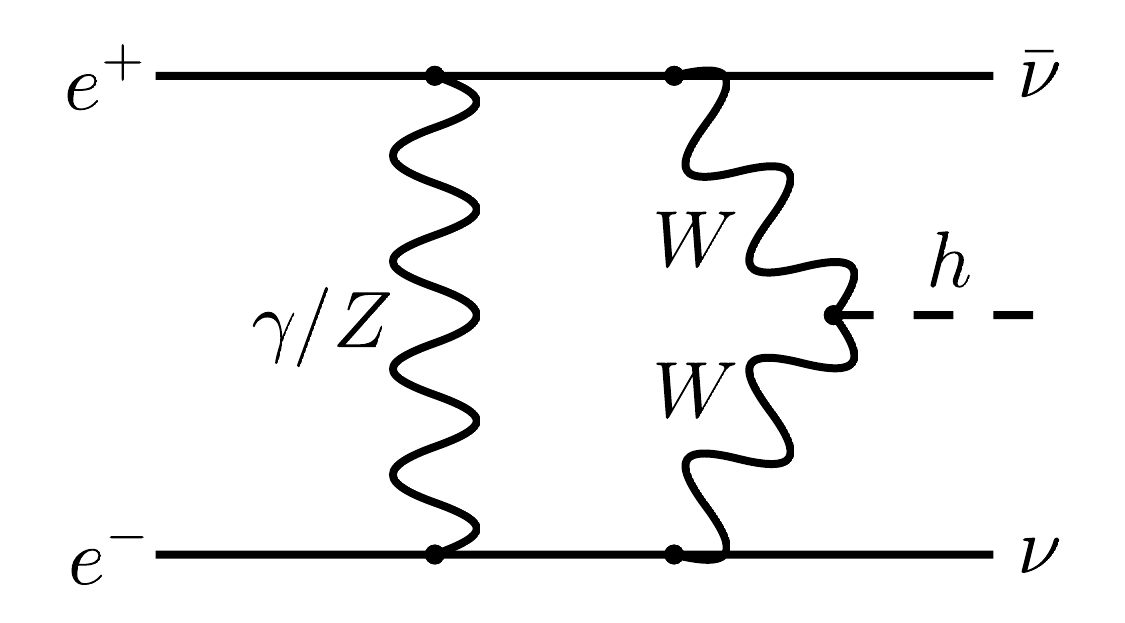} &
        \includegraphics[height=1.6cm]{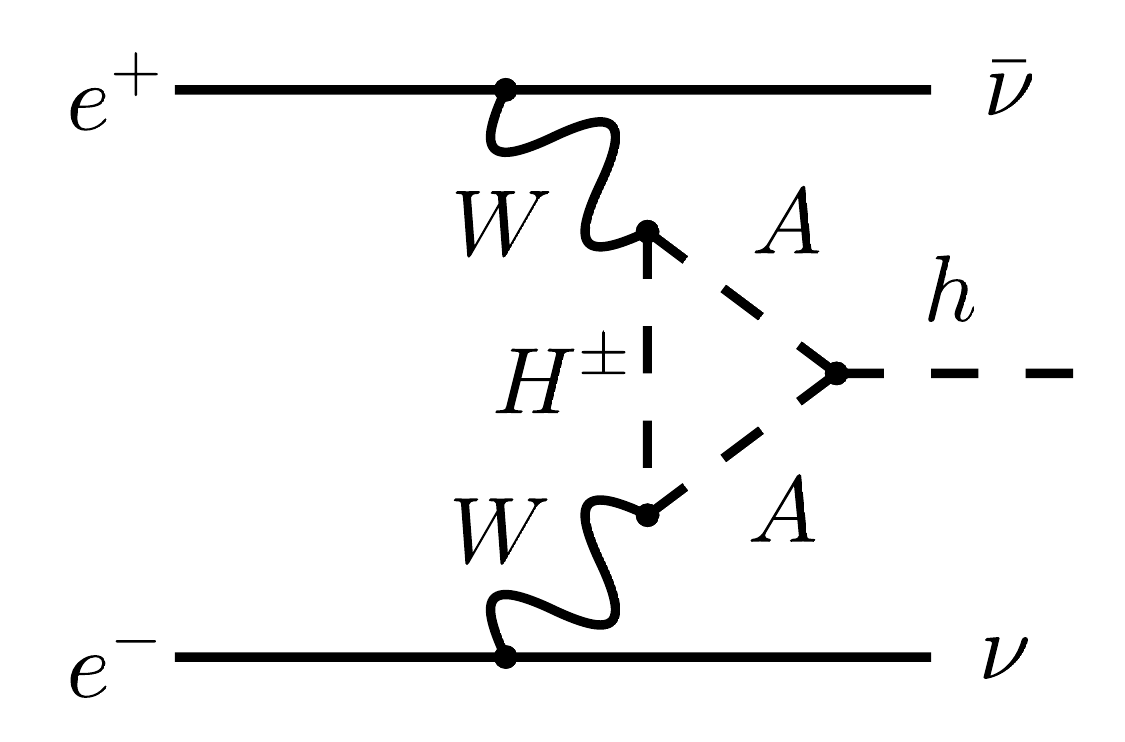} &
        \includegraphics[height=1.6cm]{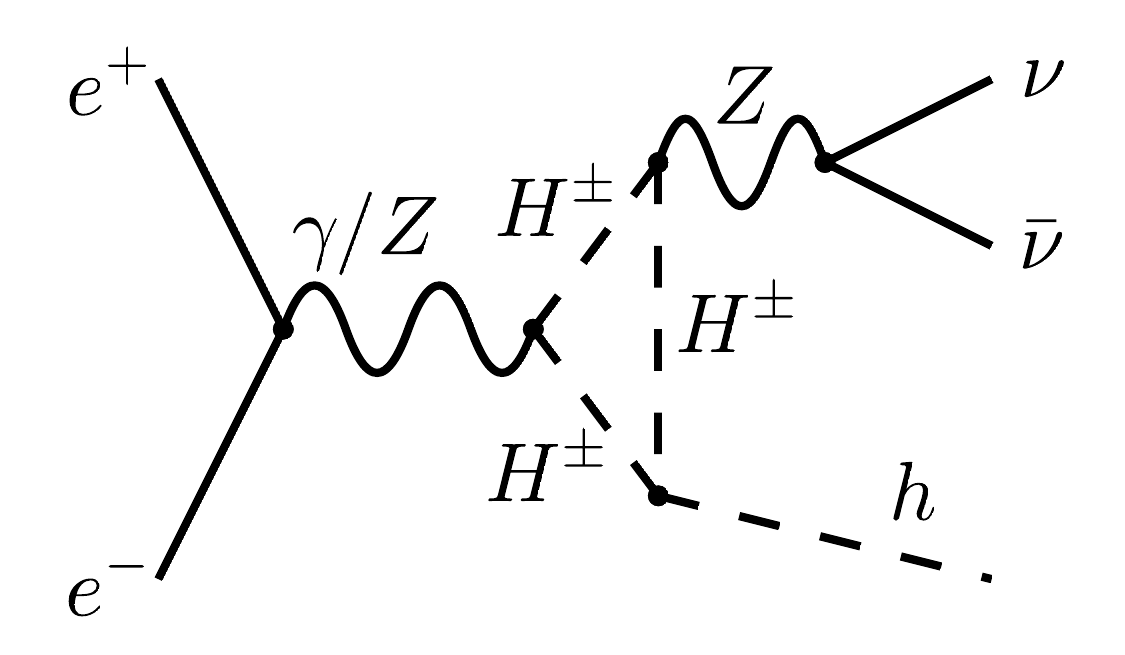} 
    \end{tabular}
        \caption{Diagrams for $e^+ e^-  \to h \, \nu \bar{\nu}$ at NLO EW in 2HDM.  
        }
        \label{fig:FeynDiag}
\end{figure}

We perform a complete NLO EW calculation in the 2HDM for $e^+ e^- \to h \, \nu \bar{\nu}$ by employing the automated NLO framework~\cite{Bredt:2022dmm,Bach:2017ggt,ChokoufeNejad:2016qux,Kalinowski:2009tq,Robens:2008sa} of the multi-purpose Monte-Carlo generator \whizard~\cite{Kilian:2007gr,Moretti:2001zz} with an interface to generic one-loop amplitude providers \texttt{OpenLoops2}~\cite{Buccioni:2019sur} and \texttt{Recola}~\cite{Actis:2016mpe, Denner:2017wsf} for the SM,
and \texttt{Recola2}~\cite{Denner:2017wsf} for the 2HDM.
We note that \texttt{OpenLoops2} supports the 2HDM at NLO QCD~\cite{Iguro:2022fel}.
Representative 2HDM Feynman diagrams are shown in Fig.~\ref{fig:FeynDiag}.  
For this study a massive electron-positron beam setup is used in \texttt{Whizard}, such that initial-state collinear singularities are regulated, while soft singularities are handled in an automated way within the Frixione-Kunszt-Signer subtraction~\cite{Frixione:1995ms}.
In the SM case, we compare our NLO EW cross section with Ref.~\cite{Denner:2003yg} and find agreement at the 0.2\% level at $\sqrt{s}=500$\,GeV. 
The small residual discrepancy can be attributed to differences in the treatment of complex masses~\cite{Denner:2003iy,Denner:2005fg,Buccioni:2019sur} and the structure function approach~\cite{Cacciari:1992pz,Skrzypek:1990qs,Skrzypek:1992vk}.
In the 2HDM case, electroweak renormalization schemes are developed in Refs.~\cite{Kanemura:2004mg,Krause:2016oke,Altenkamp:2017ldc,Denner:2018opp}. 
We employ two on-shell renormalization schemes defined in Ref.~\cite{Denner:2018opp}.
The default on-shell scheme for mixing angles in \texttt{Recola2} serves as our reference scheme, while the background-field approach is used to estimate scheme uncertainties.
$\lambda_5$ is renormalized in the $\overline{\rm MS}$ scheme.
We compare our NLO EW cross sections for $pp \to h \, \mu^+ \nu_{\mu}$ in several 2HDM benchmarks in different renormalization schemes with Ref.~\cite{Denner:2018opp} and \texttt{HAWK}~\cite{Denner:2014cla}, and we find excellent agreement at the permille level.
This is one of the most extensive NLO applications of \whizard, not due to the complexity of this process, but because of the massive scan over parameter points performed for the same process as outlined below, capitalizing on the massive parallelization features of \whizard~\cite{Brass:2018xbv,ChokoufeNejad:2014skp}.

In the following benchmark, we compare  predictions between the SM and type~I 2HDM.
The SM input parameters are
$G_{\rm F} = 1.166378\times 10^{-5}\,\text{GeV}$, $m_h = 125.2\,\text{GeV}$,
$m_Z = 91.1539\,\text{GeV}$, $\Gamma_Z = 2.4946\,\text{GeV}$,
$m_W = 80.3407\,\text{GeV}$, $\Gamma_W = 2.14\,\text{GeV}$, 
$m_b = 4.183\,\text{GeV}$, $m_t = 172.56\,\text{GeV}$,
$m_e = 5.110 \times 10^{-4}\,\text{GeV}$,
and the $G_\mu$ scheme is employed. 
The $W$ and $Z$ masses and widths are pole values, which are converted from measured on-shell values. 
Light quark and lepton mass effects are negligible for this process, except for the electron mass due to large logarithms from the initial-state radiation~(ISR).
A 2HDM benchmark point allowed by theoretical and experimental constraints based on \texttt{HiggsTools}~\cite{Bahl:2022igd} is 
$m_H = m_{H^\pm} = 400\,\text{GeV}$, $m_{A} = 435\,\text{GeV}$,
$\cba = 0.03734$, $t_\beta = 1.88$, $\lambda_5 = -2.54$.
The on-shell renormalization scheme for mixing angles is employed, and the renormalization scale for $\lambda_5$ is set to $\sqrt{s}$.

In Table~\ref{tab:2HDM} we present the LO and NLO EW total cross sections at $\sqrt{s}=365$ and $550$\,GeV (both unpolarized) in the SM and the 2HDM benchmark, as well as in the alignment limit $\cba = 0$.
At LO, increasing the energy from $\sqrt{s}=365$ to $550$\,GeV enhances the cross section by about $75\%$, both in the SM and the 2HDM. 
For both energies, we find only a permille level reduction of the cross section in the 2HDM compared to the SM, which vanishes in the alignment limit.
The situation changes at NLO: while the EW corrections are negative in all cases at the order of ten percent, they are more pronounced in the 2HDM, even in the alignment limit.
Specifically, the reduction amounts to  $1.7\%$ ($1.8\%$) and $1.9\%$ ($2.1\%$) at $\sqrt{s}=365$ ($550$)\,GeV when comparing the 2HDM cross sections, with and without alignment, to the SM.
%
Therefore, we conclude that NLO EW corrections are a very sensitive probe of BSM effects in our 2HDM benchmarks.

\begin{table}[t]
    \centering 
    \renewcommand{\arraystretch}{1.1}
    \begin{tabular}{|c|c|c|c|c|} \hline
    & \multicolumn{2}{c|}{$\sqrt{s} = 365$\,GeV} & \multicolumn{2}{c|}{$\sqrt{s} = 550$\,GeV} \\ \cline{2-5}
         & ~LO [fb]~ & NLO EW [fb]   & ~ LO [fb] ~ & NLO EW [fb] \\ \hline
        SM & 55.79 & 52.44(1) &  97.82(1)  & 88.45(2)    \\ 
        2HDM  & 55.71 & $51.45(1)$ & 97.67(1) & 86.59(2) \\ 
        \rowcolor{gray!15}
        Rel.Diff. & $-0.1\%$  & $-1.9\%$ & $-0.2\%$ & $-2.1\%$ \\ 
         2HDM & \multirow{2}{*}{55.79} & \multirow{2}{*}{51.58(1)} & \multirow{2}{*}{97.81(1)} & \multirow{2}{*}{86.83(2)} \\[-1mm]
        (aligned)  &  &   &  &\\ 
        \rowcolor{gray!15}
        Rel.Diff. & $0.0\%$ & $-1.7\%$ & $0.0\%$ & $-1.8\%$ \\ \hline
    \end{tabular}
    \caption{Total cross sections for $e^+ e^- \to h \, \nu \bar{\nu}$ in the SM and the type~I 2HDM benchmark without cuts, and the alignment limit is realized with $\cos(\beta-\alpha) = 0$.
    The relative difference $(\sigma_{\rm 2HDM}-\sigma_{\rm SM})/\sigma_{\rm SM}$ is reported at LO and NLO.
     Monte-Carlo integration errors larger than 0.01~fb are indicated in brackets.
    }
    \label{tab:2HDM}
\end{table}
%

\begin{figure}[t]
    \centering
        \includegraphics[width=1.0\linewidth]{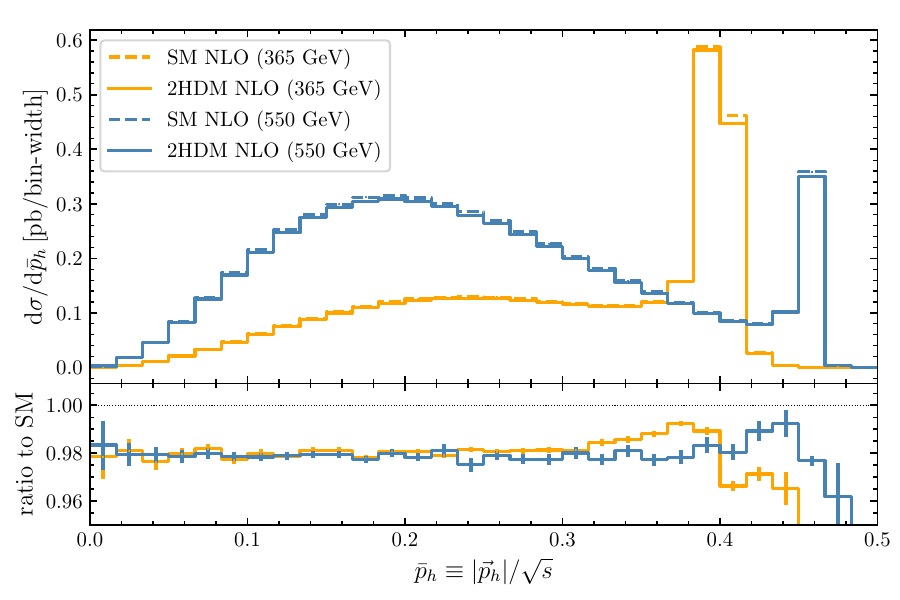}
        \caption{Top panel: differential cross sections at NLO as a function of the normalised Higgs three-momentum $\bar{p}_h\equiv|\vec{p}_h|/\sqrt{s}$ at $\sqrt{s}=365\,$GeV (orange) and $\sqrt{s}=550\,$GeV (blue), in the SM (dashed) and 2HDM (solid). 
        Bottom panel: ratios of the 2HDM predictions to the SM ones at NLO.
        }
        \label{fig:Differential}
\end{figure}

In Fig.~\ref{fig:Differential} we show the differential cross section as a function of the normalized three-momentum of the Higgs, $\bar{p}_h\equiv|\vec{p}_h|/\sqrt{s}$, again for two energies $\sqrt{s}=365$ and $550$\,GeV.
We focus on the comparison between the SM and 2HDM at NLO.
In the upper panel, for $\sqrt{s}=365\,(550)$\,GeV the peak corresponding to on-shell $Zh$ production, with the $Z$ subsequently decaying into $\nu\bar{\nu}$, is located at $\bar{p}_h\approx0.401$ ($0.459$), and values up to $\bar{p}_h\approx0.441$ ($0.474$) are kinematically allowed. 
Most of the cross section comes from the broader $WW$-fusion contribution at lower values of $\bar{p}_h$.
We observe that the relative contribution from the $Zh$ channel decreases when going to higher $\sqrt{s}$.
In the lower panel we show the ratios of the 2HDM predictions to those in the SM.
Both for $\sqrt{s}=365$ and $550\,$GeV we observe that the 2HDM effects lead to an overall reduction of the cross sections by about $2\%$, in line with the findings for the total cross sections in Table~\ref{tab:2HDM}.
For most of the $\bar{p}_h$ range the shift is flat, becoming smaller just below the $Zh$ peak, but larger once crossing to higher $\bar{p}_h$. 
The separation of $WW$ and $Zh$ channels in the differential distribution enables simultaneous probes of 2HDM effects in a single process. 
This separation could be achieved, for example, through a cut on the energy of Higgs boson and further facilitated by the use of polarized beams, which is beyond the scope of this study.

Moreover, at $\sqrt{s} = 240$\,GeV we find a similar $2\%$ reduction relative to the SM prediction (43.87~fb) at NLO.
Nevertheless, the ISR effects beyond NLO can reach several percent around 240 GeV~\cite{Denner:2003yg}, due to the kinematical effect near the $Zh$ production threshold,
we therefore focus on $\sqrt{s} = 365$ and 550\,GeV cases. 
At larger $\sqrt{s}$, our process is dominated by the $WW$-fusion channel and away from the threshold region of $Zh$ production, thus the beyond NLO ISR effects are only at the few permille level~\cite{Denner:2003yg}. 
We note that the NLO ISR effects are consistently taken into account in our calculation through the real corrections.
We also investigate the $Zh$-mediated process $e^+e^- \to h \, \mu^+\mu^-$ and find much smaller cross sections of $ 4.04\,(1.72)$~fb at $\sqrt{s} = 365\, (550)$\,GeV.
Hence, the process $e^+e^- \to h \, \nu\bar{\nu}$ is advantageous in probing new physics due to large $WW$-fusion contributions. 

We emphasize that this indirect probe at future $e^+e^-$ colliders complements direct searches for 2HDM.
At the (HL)-LHC, single production of an additional scalar via gauge interaction is suppressed by the $h$-$H$ mixing, whereas production through Yukawa interaction is highly model dependent.
Although electroweak pair production of additional scalars, e.g., 
$pp \to W^+ \to H^+ H$ is not suppressed by the mixing, the production rates decrease rapidly with increasing scalar masses.
While at $e^+e^-$ colliders, if the collision energy is sufficiently high, e.g., in the benchmark with $\sqrt{s}=550$\,GeV and $m_H=400$\,GeV, the process $e^+e^-\to Z^*\to ZH$ becomes accessible, its cross section remains suppressed by the mixing 
and vanishes in the alignment limit.
In contrast, for $e^+e^- \to h\, \nu\bar{\nu}$ we show that the difference in cross sections between the 2HDM and the SM at NLO EW can be sizable even in the alignment limit, thereby providing an important complementary window to probe the model.

In the next stage, we present a comprehensive analysis for four types of $\mathbb{Z}_2$-symmetric 2HDMs: type~I, II, X (lepton specific) and Y (flipped). 
We generate the parameter points with \texttt{ScannerS} \cite{Coimbra:2013qq,Muhlleitner:2020wwk} and \texttt{HiggsTools} \cite{Bahl:2022igd} based on \texttt{HiggsBounds} \cite{Bechtle:2008jh,Bechtle:2011sb,Bechtle:2013wla,Bechtle:2020pkv} and \texttt{HiggsSignals} \cite{Bechtle:2013xfa,Bechtle:2020uwn} to take theoretical and experimental constraints into account.
For instance, these constraints include tree-level perturbative unitarity, boundedness from below and electroweak vacuum stability as theoretical constraints, and electroweak precision observables, flavor-changing processes, direct searches at the LHC and LEP as experimental constraints. 
In our setup, the additional scalars are degenerate.
The free parameters $m_\phi \equiv m_H=m_A=m_{H^\pm}$, $\cba$, $\tb$ and $\lambda_5$ are chosen from the 95\%~confidence level~(C.L.) allowed region.
In total, we generate $30\,000$ parameter points for the type~I, II and Y, and $70\,000$ parameter points for the type~X.


\begin{figure*}[t]
    \centering
    \begin{minipage}{\textwidth}
        \includegraphics[width=1.\linewidth]{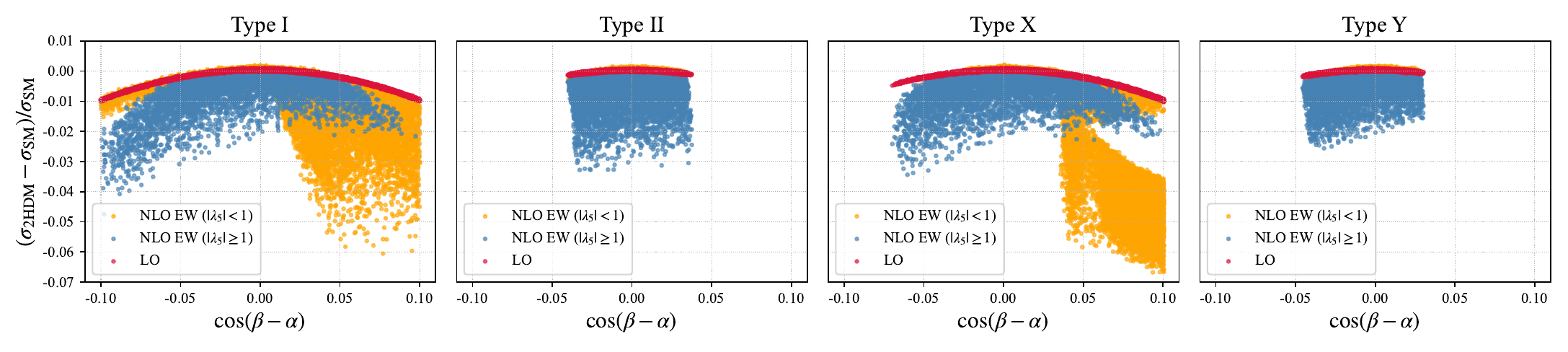} 
        \caption{Type~I, II, X and Y 2HDM predictions in the $\cos(\beta-\alpha)$ plane at $\sqrt{s} = 365$\,GeV within the allowed parameter space, with $|\cos(\beta-\alpha)|<0.1$.
    Relative differences with respect to the SM are shown at LO (red) and NLO (orange for $|\lambda_5|<1$, blue for $|\lambda_5|\geq 1$).
    The combined theoretical and experimental uncertainty is estimated to be 0.92\%.
        }
        \label{fig:cab_plots}
    \end{minipage}
\end{figure*}
\begin{figure*}[t]
    \centering
    \begin{minipage}{\textwidth}
    \includegraphics[width=1.\linewidth]{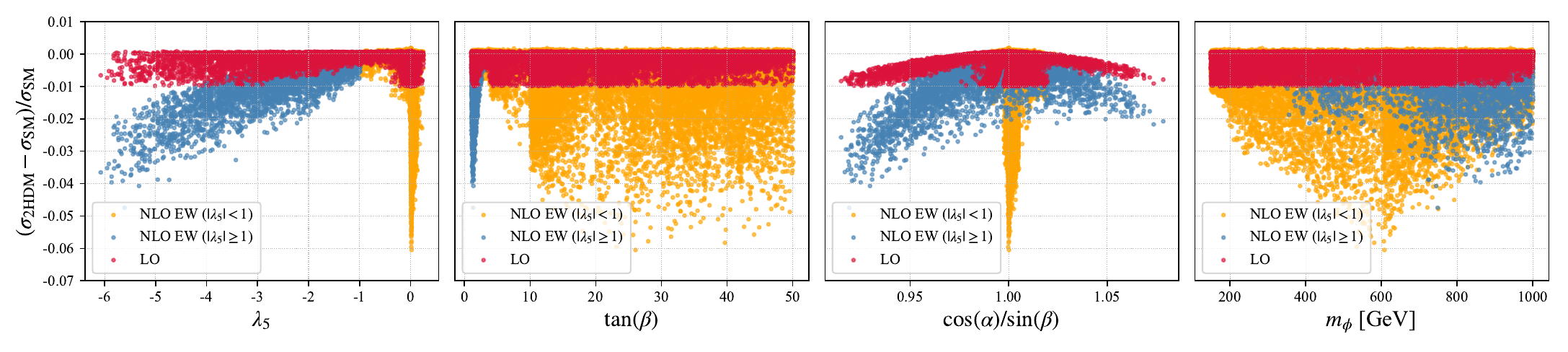} 
    \caption{Type~I 2HDM predictions in the $ \lambda_5$, $\tan(\beta)$, $\cos(\alpha)/\sin(\beta)$ and $m_{\phi} $ parameter planes at $\sqrt{s} = 365$\,GeV  within the allowed parameter space, with $|\cos(\beta-\alpha)|<0.1$.}
    \label{fig:type1-rest-365}
    \end{minipage}
\end{figure*}
\begin{figure*}[t]
    \centering
    \begin{minipage}{\textwidth}
    \includegraphics[width=1.\linewidth]{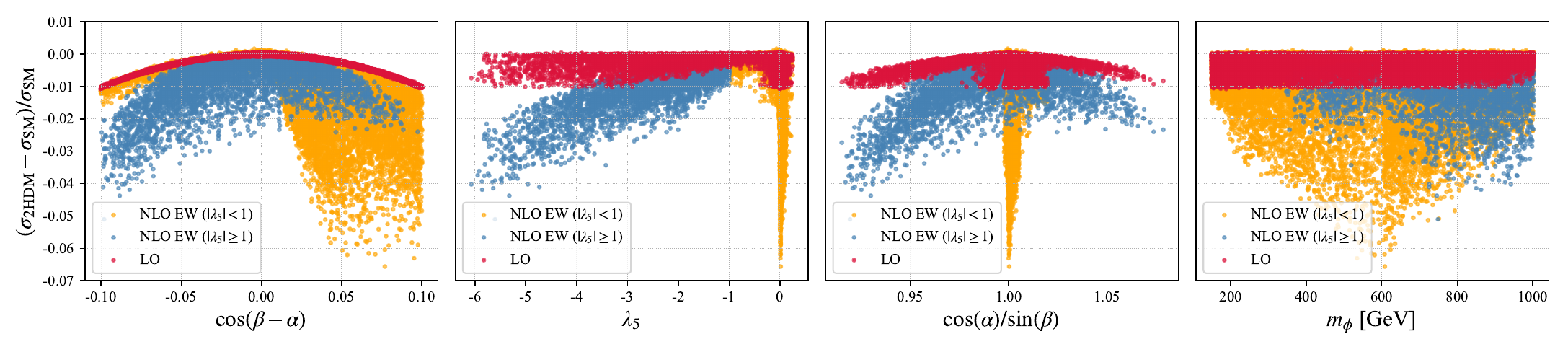} 
    \caption{Type~I 2HDM predictions at $\sqrt{s} = 550$\,GeV  within the allowed parameter space, with $|\cos(\beta-\alpha)|<0.1$. The combined theoretical and experimental uncertainties are estimated to be 0.85\%.}
    \label{fig:type1-550}
    \end{minipage}
\end{figure*}

In our sensitivity analysis, the 2HDM predictions in parameter planes of $ \cba $, $\lambda_5$, $\tb$, $\caosb \equiv \cos(\alpha)/\sin(\beta)$, $m_{\phi}$  
are presented in Figs.~\ref{fig:cab_plots} and~\ref{fig:type1-rest-365} at $\sqrt{s} = 365$\,GeV, 
and in Fig.~\ref{fig:type1-550} for $\sqrt{s} = 550$\,GeV.
The relative differences $(\sigma_{\rm 2HDM} - \sigma_{\rm SM})/\sigma_{\rm SM}$ at LO and NLO are shown for $|\cba|<0.1$, covering more than the estimated 95\%~C.L. allowed range for the least constrained type I 2HDM at future $e^+e^-$ colliders~\cite{Gu:2017ckc, Chen:2019pkq}.
To estimate the theoretical uncertainty, we compute the renormalization scheme dependence for mixing angles in the type~I 2HDM for hundreds of parameter points,
finding the maximal uncertainty of $0.7\% \,(0.8\%)$ for the relative difference at $\sqrt{s} = 365 \,(550)$\,GeV for $|\cba|<0.1$.
We note that NNLO mixed QCD-EW SM corrections yield only few permille effects in the $G_\mu$ scheme~\cite{Sun:2016bel,Wang:2021imw},
and that in the relative difference, higher-order SM corrections cancel and do not contribute to the uncertainty.
To further reduce the scheme uncertainty, NNLO EW corrections in the 2HDM would be required.
On the other hand, the experimental uncertainty of the cross section measurement will reach $0.6\%\,(0.3\%)$ at $\sqrt{s}=365\,(500)$\,GeV with $4.3\,(6.4)$\,ab$^{-1}$ of data~\cite{LinearColliderVision:2025hlt}.
Therefore, the potential gain from an NNLO-level theoretical uncertainty reduction would not affect our conclusion below.
We expect that the sensitivity difference between $\sqrt{s}=500$ and $550$\,GeV is negligible.

It is shown in Fig.~\ref{fig:cab_plots} that  $\cba$ is the key parameter governing the LO 2HDM deviation from the SM.
We first discuss the type~I 2HDM plot at $\sqrt{s} = 365$\,GeV.
The LO relative difference (red) vanishes as the $|\cba|$ approaches the alignment limit and increases with $|\cba|$ reaching at most $-1\%$. 
At NLO, however, richer phenomena arise that cannot be described by a single parameter.
We highlight the impact of the NLO EW corrections by showing two branches separating cases of small $|\lambda_5|<1$ (orange) and large  $|\lambda_5|\geq 1$ (blue).
The NLO relative difference reaches $-6\%$ in the small-$|\lambda_5|$ branch and $-4\%$ in the large-$|\lambda_5|$ branch.
The $-2\%$ deviation in the alignment limit is observed in the large-$|\lambda_5|$ branch, however, 
it does not imply that this effect is solely driven by the $\lambda_5$ self-interaction, since Fig.~\ref{fig:type1-rest-365} indicates sizable mass effects and top-Yukawa corrections deviating from the SM when $\caosb \neq 1$. 
Note that the ratio $\caosb$ governs the top-Yukawa coupling to the Higgs in the same way across all four 2HDM types~\cite{Branco:2011iw}.
We find that the relative difference at $\sqrt{s}=550$\,GeV in Fig.~\ref{fig:type1-550} is similar to $\sqrt{s}=365$\,GeV but slightly more pronounced. 
Since the expected accuracy of the cross section measurement will be better at higher energies,
we conclude that the operation at $550$\,GeV will be more sensitive to the 2HDM effects.
For the remaining plots in Fig.~\ref{fig:cab_plots}, the relative differences reach about $-7\%$ in the type~X, and around $-3\%$ in the more constrained type~II and Y.
We note that although our NLO predictions for Higgs production are insensitive to different 2HDM Yukawa structures,
the allowed parameter space across types determine the shapes of distributions, providing restricted ability for the 2HDM-type discrimination.
To further distinguish them, we need to include 
Higgs decay channels.
We note that since both the beyond NLO ISR effects and the NNLO mixed QCD-EW corrections are at the few permille level, the discrimination between the SM and the 2HDMs observed at NLO EW is valid even without a full NNLO calculation.

To scrutinize the NLO effects, the relative differences in other parameter planes at $\sqrt{s} = 365$\,GeV are shown in Fig.~\ref{fig:type1-rest-365} for the type~I 2HDM.
Although the large-$|\lambda_5|$ branch naturally develops sizable effects, the small-$|\lambda_5|$ branch 
localized near $\lambda_5 \approx 0$ and $\caosb \approx 1$, yet spread across $m_\phi$ from 200~GeV to 1~TeV,
is particularly interesting.
This implies that these large deviations are genuine NLO effects, with mixing angles and $m_\phi$ all playing a role. 
In addition, there is no clear correlation between the NLO effects and the parameters $\tb$ and $m_\phi$.
This phenomenon cannot be easily parameterized within an effective field theory approach, highlighting the importance of NLO calculations in UV-complete BSM theories.

In summary, we conduct the first full NLO EW study of Higgs production $e^+e^-\to h \,\nu \ov \nu$ in all four types of $\mathbb{Z}_2$-symmetric 2HDMs, over a vast allowed parameter space.
We find that the 2HDM effects are significantly enhanced at NLO, with deviations from the SM predictions reaching $-6\%$ to $-7\%$ with $|\cba|<0.1 $ for $\sqrt{s} = 365$ and $550$\,GeV.
Even in the Higgs alignment limit, these deviations can reach $-2\%$ to $-3\%$, making them experimentally observable.
We emphasize that this discrimination power between the SM and 2HDMs arises only at NLO, but is absent at LO.
We show that the differential distributions disentangle the $WW$-fusion and $Zh$ channels, allowing simultaneous new physics probes.
In a broader context of precision physics at future $e^+e^-$ colliders,
our findings underscore the crucial importance of higher-order electroweak calculations in UV-complete BSM theories for new physics searches,
opening up new windows that are complementary to the LHC searches.

\vskip 0.3cm
\noindent\textbf{Acknowledgment:}
We thank Georg Weiglein, Johannes Braathen and Jean-Nicolas Lang for useful discussions, and thank Matthias Steinhauser and Gudrun Heinrich for useful comments on the manuscript.
H.~Zhang is funded by the European Union under the Marie Sk{\l}odowska-Curie Actions (MSCA) grant 101202083 -- ``HINOVA", and by the Swiss National Science Foundation (SNSF) grant TMSGI2 211209.
P.~Bredt, M.~H\"ofer and W.~Kilian are supported by the Deutsche Forschungsgemeinschaft (DFG, German Research
Foundation) under grant 396021762 -- TRR 257 ``Particle Physics Phenomenology after the Higgs Discovery".
J. Reuter is supported by the DFG under the German Excellence Strategy-EXC 2121 “Quantum Universe”-390833306, and the National Science Centre (Poland)
under OPUS research project no. 2021/43/B/ST2/01778.
Y.~Ma is supported by a Postdoctoral Fellowship of the Fond de la Recherche Scientifique de Belgique (F.R.S.-FNRS), Belgium, and the IISN-FNRS convention 4.4517.08 ``Theory of fundamental interactions".
T.~Banno is supported by the JST SPRING grant JPMJSP2125.
S.~Iguro is supported by the JSPS KAKENHI grants 22K21347, 24K22879, 24K23939, 25K17385, JPJSCCA20200002 and the Toyoaki scholarship foundation.
S.~Iguro and H.~Zhang also thank the Paul Scherrer Institute and the University of Zurich for their hospitality during the course of this project.



\bibliographystyle{utphys}
\bibliography{ref.bib}

\end{document}